\documentclass[graybox]{svmult}

\usepackage{mathptmx}       
\usepackage{helvet}         
\usepackage{courier}        
\usepackage{type1cm}        
\usepackage{makeidx}         
\usepackage{graphicx}        
\usepackage{multicol}        
\usepackage[bottom]{footmisc}

\makeindex             

\begin{document}

\title*{Digital Sky Surveys from the ground: Status and Perspectives}
\titlerunning{Digital Sky Surveys} 
\author{Tom Shanks}
\institute{Tom Shanks \at Dept. of Physics, Durham University, England. \email{tom.shanks@durham.ac.uk}
}
%
%
\maketitle

\abstract*{I first review the status of Digital Sky Surveys. The focus
will be on extragalactic surveys with an area of more than 100deg$^2$.
The Sloan Digital Sky Survey is the archetype of such imaging surveys
and it is its great success that has prompted great activity in this
field. The latest surveys explore  wider, fainter and  higher resolution
and also a longer wavelength range than SDSS. Many of these surveys
overlap particularly in the S Hemisphere where we now have Pan-STARRS,
DES and the ESO VST  surveys, and our aim here is to compare their
properties. Since there is no  dedicated article on the VST ATLAS in this
symposium, we shall especially review the properties of this particular
survey. This easily fits onto our other main focus which is to compare
overlapping Southern Surveys and see how they best fit with the
available NIR imaging data. We conclude that the Southern Hemisphere
will soon overtake the North in terms of multiwavelength imaging.
However, we note that the South has more limited opportunities for  
spectroscopic follow-up and this weakness will persist during the LSST
era. Some new perspectives are offered on this and other aspects of 
survey astronomy.}

\abstract{I first review the status of Digital Sky Surveys. The focus
will be on extragalactic surveys with an area of more than 100deg$^2$.
The Sloan Digital Sky Survey is the archetype of such imaging surveys
and it is its great success that has prompted great activity in this
field. The latest surveys explore  wider, fainter and  higher resolution
and also a longer wavelength range than SDSS. Many of these surveys
overlap particularly in the S Hemisphere where we now have Pan-STARRS,
DES and the ESO VST  surveys, and our aim here is to compare their
properties. Since there is no  dedicated article on the VST ATLAS in this
symposium, we shall especially review the properties of this particular
survey. This easily fits onto our other main focus which is to compare
overlapping Southern Surveys and see how they best fit with the
available NIR imaging data. We conclude that the Southern Hemisphere
will soon overtake the North in terms of multiwavelength imaging.
However, we note that the South has more limited opportunities for  
spectroscopic follow-up and this weakness will persist during the LSST
era. Some new perspectives are offered on this and other aspects of 
survey astronomy.}

\section{Introduction}
\label{sec1}
Digital imaging sky surveys are one of the most powerful tools for the
modern astronomer. Their scientific heritage includes the 20th Century
Schmidt photographic surveys that surveyed the sky from Palomar, Siding
Springs and La Silla to $g\approx21$m in blue and red bands (e.g. Cannon
et al, 1984, Woltjer et al 1984). It took some time for CCD imaging
cameras to catch up in terms of field-of-view/area with the photographic
surveys but by the beginning of the 21st Century they had. The success
of the Sloan Digital Sky Survey (Stoughton et al., 2002) has been
enormous and the new generation of sky surveys can only hope to emulate
its success.

\begin{table}
\centering
\begin{tabular}{cccccccc}
\hline\hline
Survey & Type & Epoch & Bands & Lim. & deg$^2$ & N/S & Seeing \\
       &      &       &        & Mag. &        &     & (arcsec) \\

\hline\hline

DENIS & NIR & 1997-03 & iJK & $K\approx12$&20000&South&3 \\
SDSS & Visible & 2000-05 & $ugriz$ & $r\approx22.7$&14500 & North&1.2 \\
CFHT RCS2& Visible & 2002-09& $grz$ & $r\approx24.8$ & 830 & N$+$S&0.9\\
CFHTLS Wide &Visible & 2003-12& $ugriz$ &$r\approx25$ & 157& North&0.9\\
2MASS & NIR & 1997-01& $JHK$ & $K\approx14.3$& All sky& N+S&1.5\\
UKIDSS& NIR& 2005-12& $YJHK$ &$K\approx18.4$ &7500 &North &0.9\\
WISE  & Mid-IR & 2010-12& $3.4-22\mu m$ &$W1\approx17$ & All Sky & N$+$S & 6\\
Pan-Starrs 3$\pi$& Visible & 2010-14& $grizy$& $r\approx22.8$&30000 & N$+$S &1.1\\
SkyMapper &Visible & 2009-& $uvgriz$ &$r\approx22.0$ & 20000& South &2.5\\
VST ATLAS & Visible & 2011- & $ugriz$ & $r\approx22.7$&4700 & South&0.9\\
VST KiDS & Visible& 2011- & $ugri$& $r\approx24.6$ & 1500&South&0.7\\
VISTA VHS & NIR & 2010-& $YJK_s$&$K_s\approx18.4$ & 18000 & South& 0.7 \\
VIKING & NIR & 2010-& $zYJHK_s$ & $K_s\approx19.5$& 1500 & South &0.9\\
DES & Visible& 2013- & $grizy$& $r\approx25.0$& 5000 &South &0.9\\
DECaLS & Visible & 2015- & $grz$& $r\approx23.6$& 9000 &North &1.2\\
HSC Wide & Visible& 2015- & $grizy$& $r\approx26.0$& 1400 &South &0.7\\
\hline
\end{tabular}
\caption[]{Recent Optical and NIR extragalactic imaging  sky  surveys with 
an area of $>100$deg$^2$. Magnitude limits are quoted in $r_{AB}$
and $K_{Vega}$.
}
 \label{tb:surveys}
\end{table}

Table 1 shows the list of extragalactic imaging surveys of area
$>100$deg$^2$ that have been done until now from their starting point
around the year 2000. They start with the DENIS NIR Survey from a 1-m
telescope covering the S Hemisphere to $K\approx12$ and SDSS covering
the N Hemisphere in the optical to $r\approx22.7$ from a 2.5-m telescope
at Apache Point, New Mexico. The CFHT MegaCAM instrument also featured
powerfully in the early digital surveys with its 1deg field and high UV
sensitivity producing the Red Cluster Sequence surveys to $r\approx24.8$
over 900deg$^2$ and then the CFHT Legacy Survey with its Wide component
extending over some 157deg$^2$. NIR detectors were increasing in size,
leading then to the 2MASS survey in JHK over the full sky and UKIDSS
much deeper in YJHK but with only partial coverage of the northern sky.
Also completed by 2012 was the WISE survey in the 3.6, 4.5 12 and 22
micron bands. This was, of course,  a satellite survey but as we shall
see this meshes in with the ground-based optical surveys that it is hard
to leave out of this list. This leads to the more modern surveys that we
are going to feature specifically in this review. These include
PanSTARRS 3$\pi$ survey which covers 30000deg$^2$ in $grizy$
predominantly in the North but also reaches as far South as Dec=-30 in
the Southern Hemisphere. Then there are the ESO Southern Hemisphere
surveys using the VLT Survey Telescope in the optical for KiDS and ATLAS
and using the VISTA telescope in the NIR for the VISTA Hemisphere Survey
(VHS) and for VIKING. Finally there is the DES and DECaLS surveys
imaging in  respectively $grizy$ to $r\approx25.2$ over 5000deg$^2$ in
the Southern Hemisphere and $grz$ to $r\approx23.6$ in the Southern half 
of the SDSS footprint.

Currently it is becoming apparent that the greatest digital imaging
strides are currently being taken in the Southern Hemisphere due to the
availability of VISTA, DECam and VST OmegaCam. The focus of this review
is to assess the complementarity of the  Southern optical and NIR
imaging surveys. The case that will be made is that DES is a powerful
optical survey but can still be matched by combinations of the other
surveys such as KiDS, Pan-STARRS and ATLAS. Also DES may be less well matched
than the others in terms of the depth of its accompanying NIR surveys.

%
\begin{figure}[t]
\includegraphics[scale=.25]{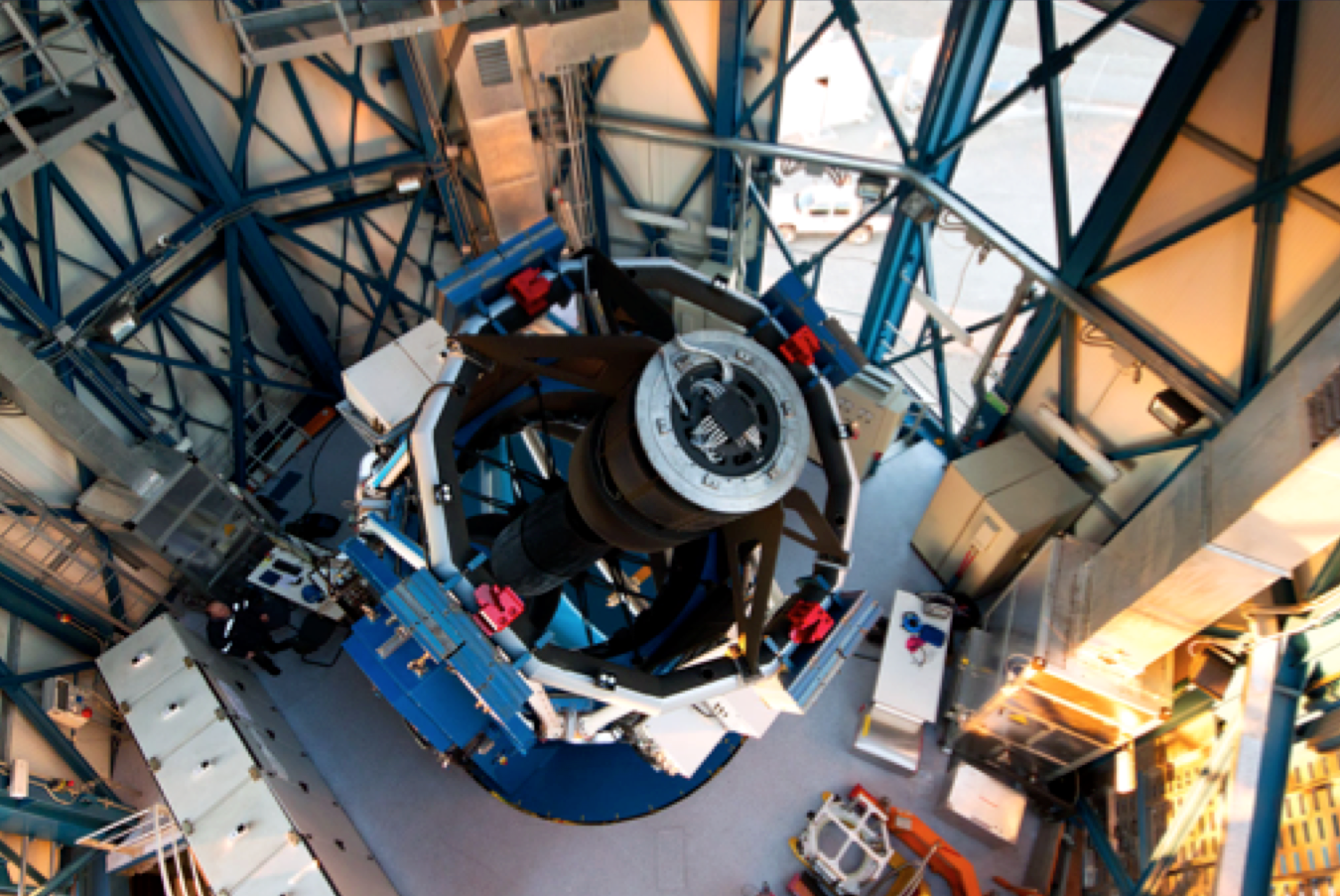}
%
%
\caption{The 2.61-m ESO VLT Survey Telescope (VST) in its dome at Paranal Observatory.}
\label{fig:vst}       
\end{figure}

\section{ESO Surveys}
\label{sec_eso}

Here we call the `ESO surveys' the surveys by the ESO VST (Schipani et
al 2012, see Fig. 1) and VISTA (Sutherland et al 2015) telescopes. We
shall discuss the ATLAS (Shanks et al 2013, 2015) and  KiDS (de Jong et
al 2013) surveys from VST and the VHS (McMahon et al 2013) and VIKING
(Sutherland et al 2012) surveys from VISTA. We shall discuss the VST
ATLAS survey at greater length since there is no dedicated ATLAS article
elsewhere in this volume. The sky footprints of the ESO ATLAS, KiDS,
VHS, VIKING and PanSTARRS $3\pi$, DES and PanSTARRS surveys are shown in
Fig. 2.

\begin{figure}[t]
\includegraphics[scale=.4]{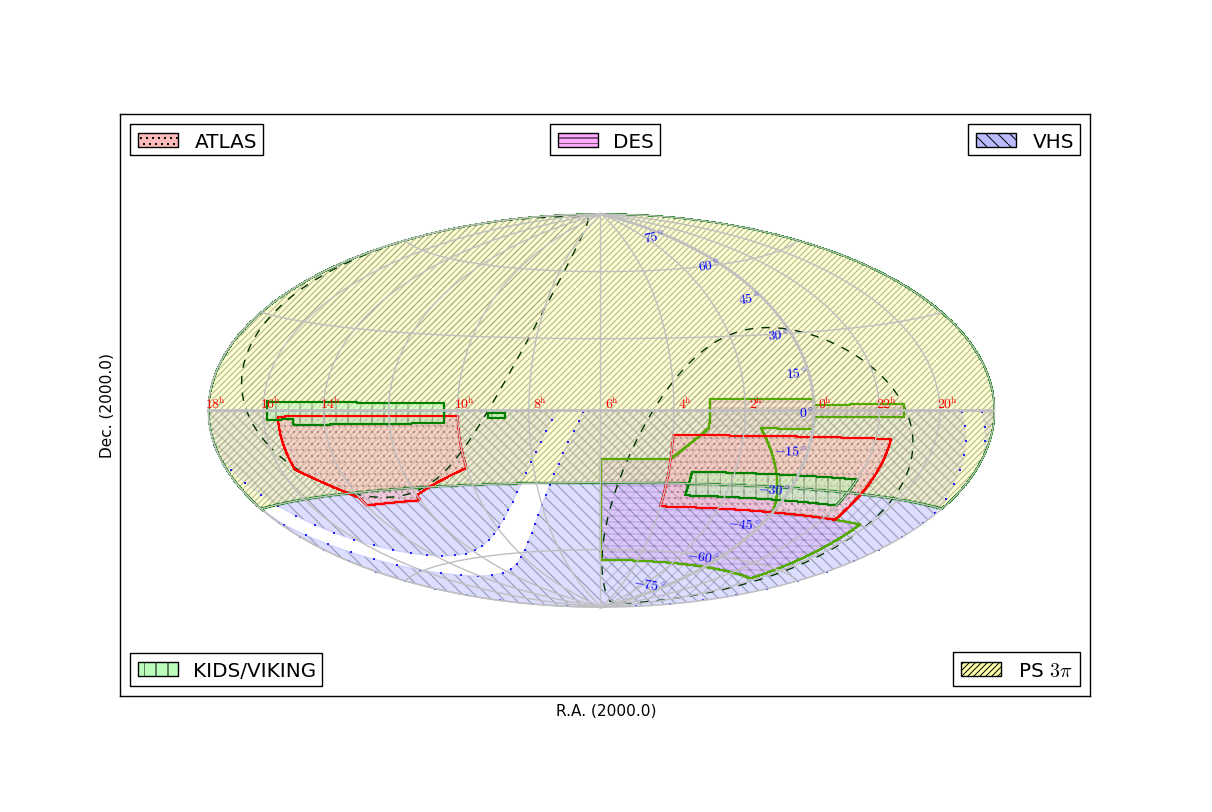}
\caption{The sky areas covered by the  ESO VST ATLAS, KiDS, ESO VISTA VHS, VIKING, DES and PanSTARRS
$3\pi$ surveys.}
\label{fig:footprints}       
\end{figure}

\subsection{VST ATLAS}
\label{subsec_atlas}

The ESO VST ATLAS targets 4700deg$^2$ of the Southern Hemisphere in the
$ugriz$ bands to similar depths to the  SDSS survey in the North. Shanks
et al (2015) report that the throughput for VST $+$ OmegaCam is usually
similar to SDSS but the seeing for VST in all bands is significantly
better than for SDSS. For example, ATLAS achieves a median of
$\approx0.''8$ FWHM in the $i$ band   and $\approx1.''0$ in the $u$
band; the equivalent numbers for SDSS are $1.''2$ and $\approx 1.''5$ so
in general a  50\% improvement. This means there is generally an
advantage for ATLAS over SDSS in terms of the $5\sigma$ magnitude limit
for stellar sources amounting to $\approx0.2$mag in $ugr$ and rising to
$\approx0.3$mag in $i$ and $\approx0.7$mag in $z$ where the throughput is also
50\% higher than for SDSS. For resolved galaxies the ATLAS and SDSS
magnitude limits are more comparable.  All of the ATLAS $ugriz$ bands have 
$\approx2\times$ longer exposure times than the $54s$ of SDSS to
accommodate the smaller $0.''2$ pixel size of OmegaCam to account for
the increased readout noise and for the use of grey time in the $iz$ bands.
There is also an ongoing `Chilean ATLAS $u$-band extension' (PI L. Infante) 
where the ATLAS $u$-band exposure time is doubled to 4 minutes.

\begin{figure}[t]
\centering{
\includegraphics[scale=.18]{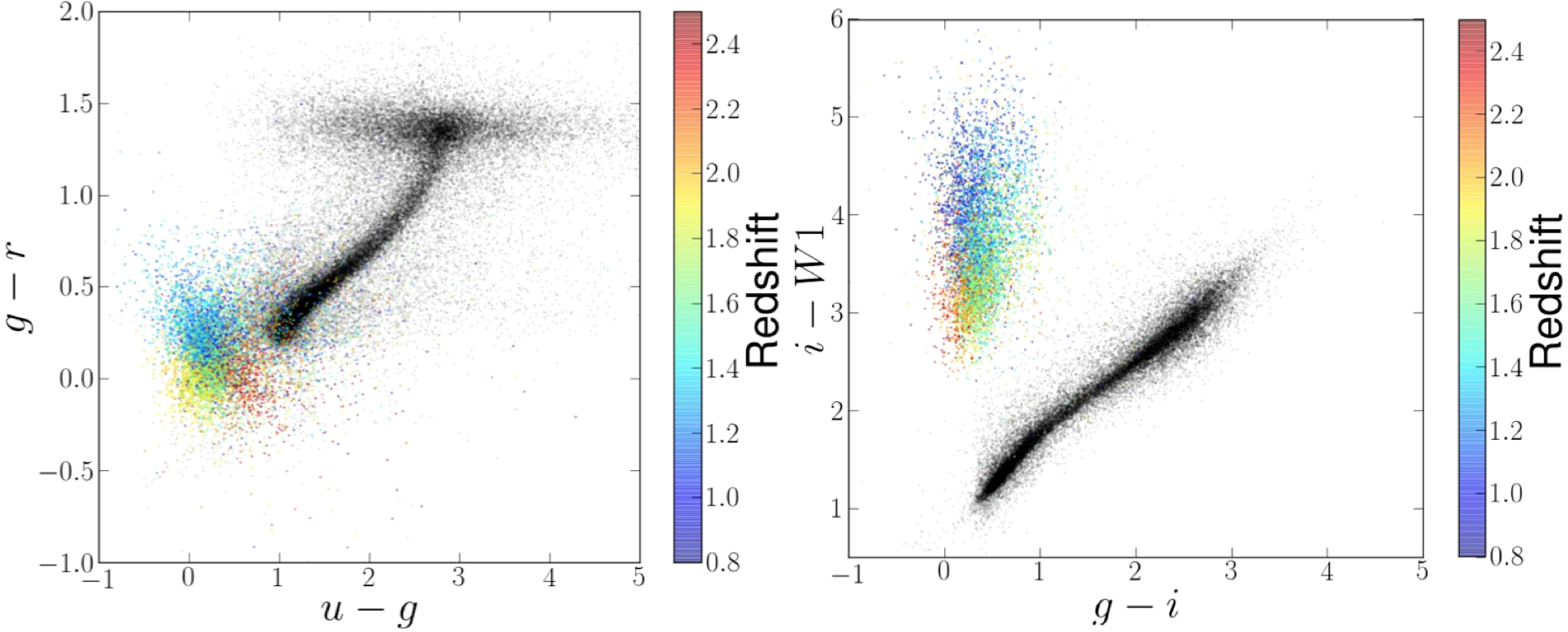}
\caption{Left: $ugr$ quasar selection.    Right: $giW1$ quasar
selection. In an ATLAS-based survey of $\approx10000$ quasars, Chehade
et al (2015 in prep). find that although the $giW1$ separates quasars from
stars more cleanly, $ugr$ selection still reaches fainter limits,
$g\approx22.5$mag}
}
\label{fig:atlas_qso}       
\end{figure}

\begin{figure}[t]
\centering{
\includegraphics[scale=.2]{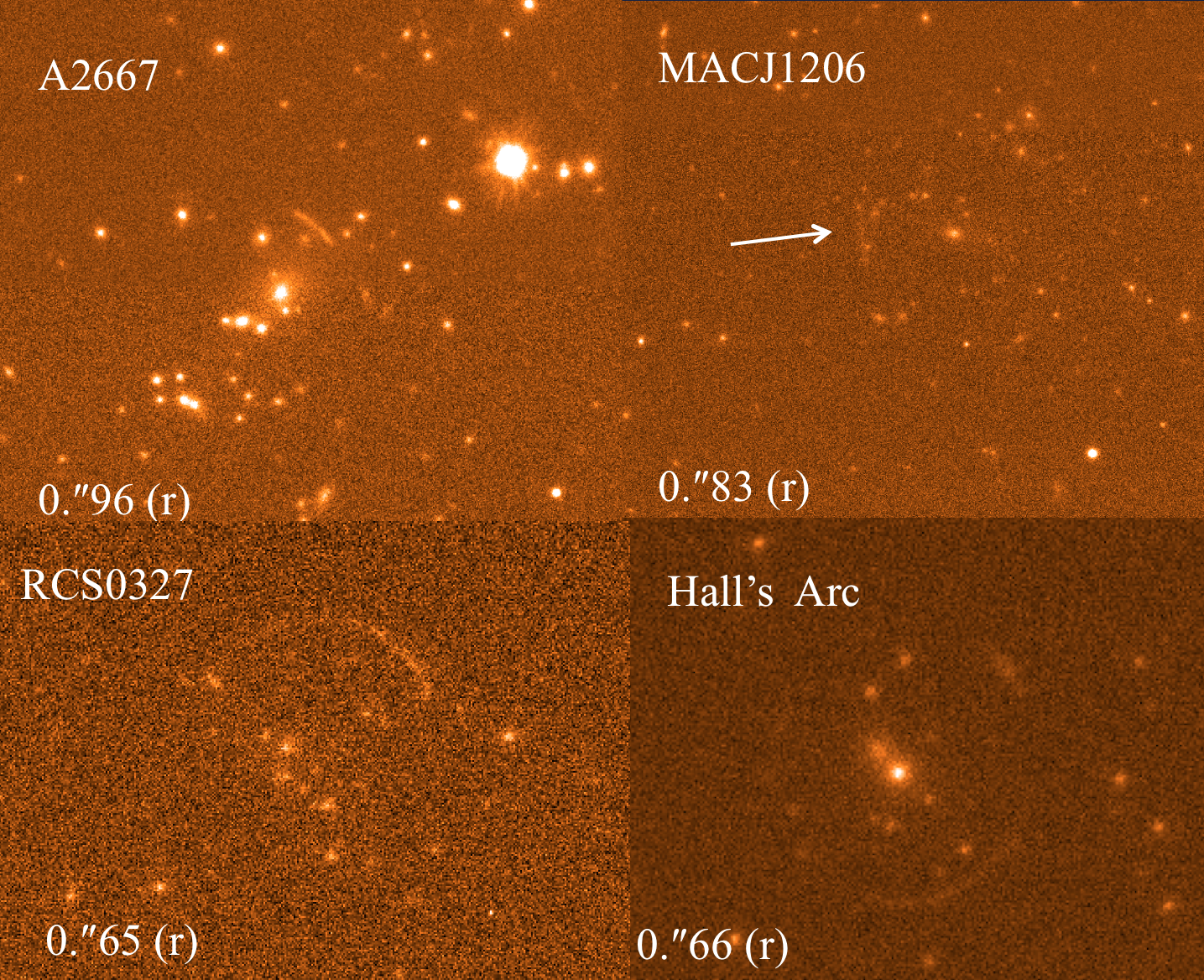}
\caption{Previously known galaxy clusters with  gravitationally lensed 
arcs detected in subarcsecond ATLAS images.}
}
\label{fig:vst_arcs}       
\end{figure}

The original science aims of ATLAS were first to exploit the $u$-band
coverage  of ATLAS to make UVX quasar searches. Now one such survey has
already been done, the 2QDES pilot survey, comprising $\approx10000$
quasar redshifts to $g\approx22.5$ (Chehade et al 2015 in prep.). We
found that $u$ selection still reaches deeper than cuts that use the
WISE W1, W2, such as $g-i:i-W1$, although these appear to show less
contamination by stars (see Fig. 3). Another top priority for VST ATLAS
is to search for the Integrated Sachs-Wolfe effect in the Southern Hemisphere
by cross-correlating the positions of Luminous Red Galaxies (LRGs)  and
microwave background fluctuations. Here again we shall be selecting LRGs
using traditional $griz$ colour cuts complemented by WISE magnitudes at
higher redshifts. Photo-z catalogues  for ATLAS are being prepared and
we find the best results using ANNz neural networks that incorporate
both ATLAS $ugriz$ and WISE $W1$, $W2$ magnitudes.

The improved seeing also opens up new avenues for ATLAS projects in
terms of both finding multiple quasar lenses and lensed arcs in galaxy
clusters (see Fig. 4). Most ATLAS tiles have at least one band with
significantly sub-arcsecond seeing. P.L. Schechter has therefore 
initiated a programme to find quadruple lenses in ATLAS images.
Meanwhile Carnall et al (2015) have combined ATLAS and WISE photometry
to confirm spectroscopically confirm 3 bright $z>6$ quasars, the first
from ATLAS.

\begin{figure}[t]
\centering{
\includegraphics[scale=.3]{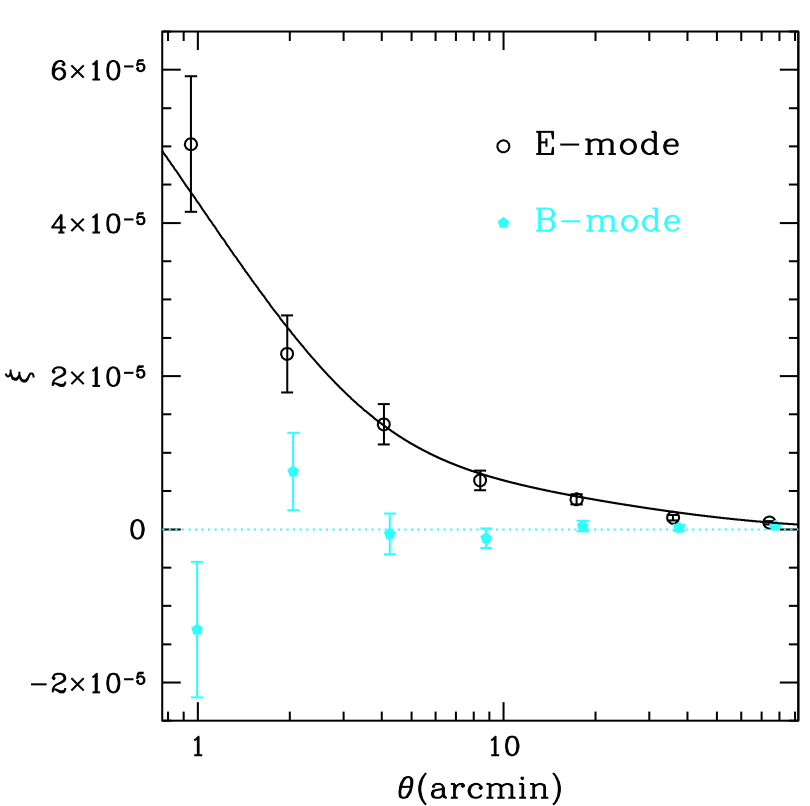}
\caption{Weak lensing shear results from KiDS, measuring the mass
correlation function at $z\approx0.4$. The low amplitude of the B-modes
provides upper limits on how  systematics affect  the lensing E-modes.}
}
\label{fig:kids_lensing}       
\end{figure}

\subsection{VST KiDS}
\label{subsec_kids}
KiDS (PI K Kuijken) is the premier VST survey, aimed at covering
1500deg$^2$ in $ugri$ split between an NGC field at the equator and an
SGC field at Dec=-30. These field comprise the original areas of the 2dF
Galaxy Redshift Survey. The survey aims to reach $r\approx25.2$ with the
best seeing on the VST reserved for the KiDS $r$ band where the median
seeing is currently $\approx 0.''7$. The main purpose is to do weak
lensing tomography and, with the remarkably distortion-free field of VST
and 9-band photometry (including VISTA VIKING $zYJHK$) for photo-z, KiDS
seems ideally positioned. KiDS DR2 released 172deg$^2$ with another
year's worth of data already taken (de Jong et al. 2013) The DR2
coverage means KiDS' area is already larger than the CFHTLS Wide survey,
the previously largest lensing survey. The main problem for KIDS is that
currently it is estimated to take until 2020 to complete. The $i$ band
is already reasonably complete but $ugr$ are further behind. Fig. 5
shows weak lensing results from KiDS that measure the mass correlation
function at $z\approx0.4$ (K. Kuijken, priv. comm.).

\subsection{VISTA VHS}
\label{subsec_vhs}
VISTA VHS (McMahon et al 2013) is the sister survey to VST ATLAS and
also to DES. VISTA is a 4.1-m telescope sited near Cerro Paranal and
dedicated to NIR imaging. Its 1.5deg$^2$ tile `images' are composed of 6
individual pawprints. VHS-ATLAS covers 5000deg$^2$ mainly overlapping
the VST ATLAS survey with exposures of $Y$(120s), $J$(60s), $K_s$(60s). VHS-DES
covers 4500deg$^2$ (excluding the 500deg$^2$ of the SGC VIKING
footprint) and complements the DES survey with  exposures of $J$(120s) and
$K_s$(120s). At lower latitudes there is also VHS-Galactic Plane survey
covering another 8200deg$^2$ with exposures of J(60s), K(60s). Thus in
the ATLAS area VHS reaches $K_s=18.2$(Vega), similar to the UKIDSS limit,
and in the DES area it  reaches $K_s=18.5$. Here again VHS takes the
poorer seeing VISTA time with the better seeing going to VIKING and
other projects such as UltraVISTA. One of the prime aims of VHS is to
look for high redshift quasars and therefore that is why there is so
much emphasis on $Y$, $J$. (In the DES area $Y$ is done by  DECam). Other aims
are to detect the faintest brown dwarf stars and to provide NIR
magnitudes to help measure  photo-z for the ATLAS and DES surveys.
Currently VHS ATLAS is $\approx50$\% complete and VHS-DES is
$\approx70$\% complete.

\subsection{VISTA VIKING}
\label{subsec_viking}
The VIKING survey (Sutherland et al 2012) aims to cover the same 1500deg$^2$ area as KiDS in
$zYJHK_s$ to a 5$\sigma$ detection limit of $K_s\approx19.5$ (Vega). VIKING
includes the 2dF and GAMA galaxy redshift survey areas, the Herschel
ATLAS, WALLABY-ASKAP HI survey and the 2dFLens survey. VIKING is now
back on the sky again after a gap of about a year due to management
changes. VIKING has now covered  $\approx 1000$deg$^2$ with at least one
of $ZYJ$ or $JHK_s$ visits. This is a good match to the KiDS coverage in
the $i$-band.

\begin{figure}[t]
\includegraphics[scale=.35]{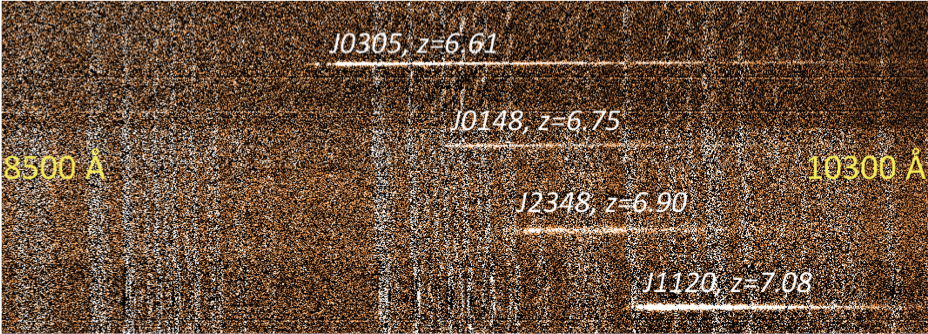}
\caption{3 VIKING $z>6$ quasars compared to the $z\approx7$ quasar from UKIDSS.}
\label{fig:viking_ukidss_2mass}       
\end{figure}

Scientifically, the survey has already found several high redshift $z>6$
quasars (Venemans et al 2013 - see Fig. 6). At the opposite extreme of
redshifts, the additional depth of VIKING over UKIDSS LAS or the VHS
means that galaxy surface brightness sensitivity is significantly better
(see Fig. 7). So for brighter galaxies at $z<0.2$,  NIR morphological
parameters can be extracted and this is a primary focus of the GAMA
survey.

\begin{figure}[t]
\includegraphics[scale=.5]{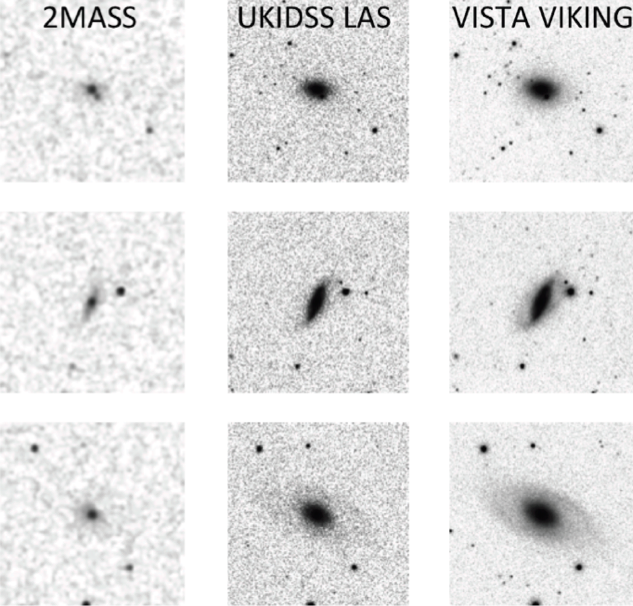}
\caption{Left: 2MASS Middle: UKIDSS  Right: VIKING. These K-band  images
of 3 galaxies shows the significantly fainter  surface brightness
detection limits of VIKING as compared to 2MASS and UKIDSS.}
\label{fig:viking_ukidss_2mass}       
\end{figure}

The parameters extracted from the imaging (total stellar mass,
bulge-to-disk radio, etc) are a key component of the GAMA project (PIs
Driver and Hopkins) that covers a significant part of the VIKING/KiDS
area with AAOmega to recover spectra for a much denser sampling than
traditional redshift surveys. GAMA will be complemented with the WALLABY radio survey 
that will recover the 21cm HI emission of $z<0.25$ galaxies.

The largest Open Time Key Projects with Herschel was H-ATLAS (PI Eales).
It covers $\approx550$deg$^2$ of which $\approx 400$deg$^2$ is in the
VIKING/KiDS footprint. The wide redshift range of the Herschel sources, accentuated by lensing, 
means that NIR imaging is crucial to the identification of these sources.

The VIKING/KiDS NGC area will also be  covered by the HyperSuprimeCam
(Miyazaki et al 2012) Wide survey area (2015-20) which will image
$grizY$ to 26.5, 26.1, 25.9, 25.1 and 24.4, $\approx2.5$. mag fainter
than the KiDS limits and $\approx1$ mag fainter than the DES 5 Year
limits (see Table \ref{tb:exp_time}). In total HSC Wide will cover 1400deg$^2$ 
at Dec$\approx0$deg.

\begin{figure}[t]
\includegraphics[scale=.17]{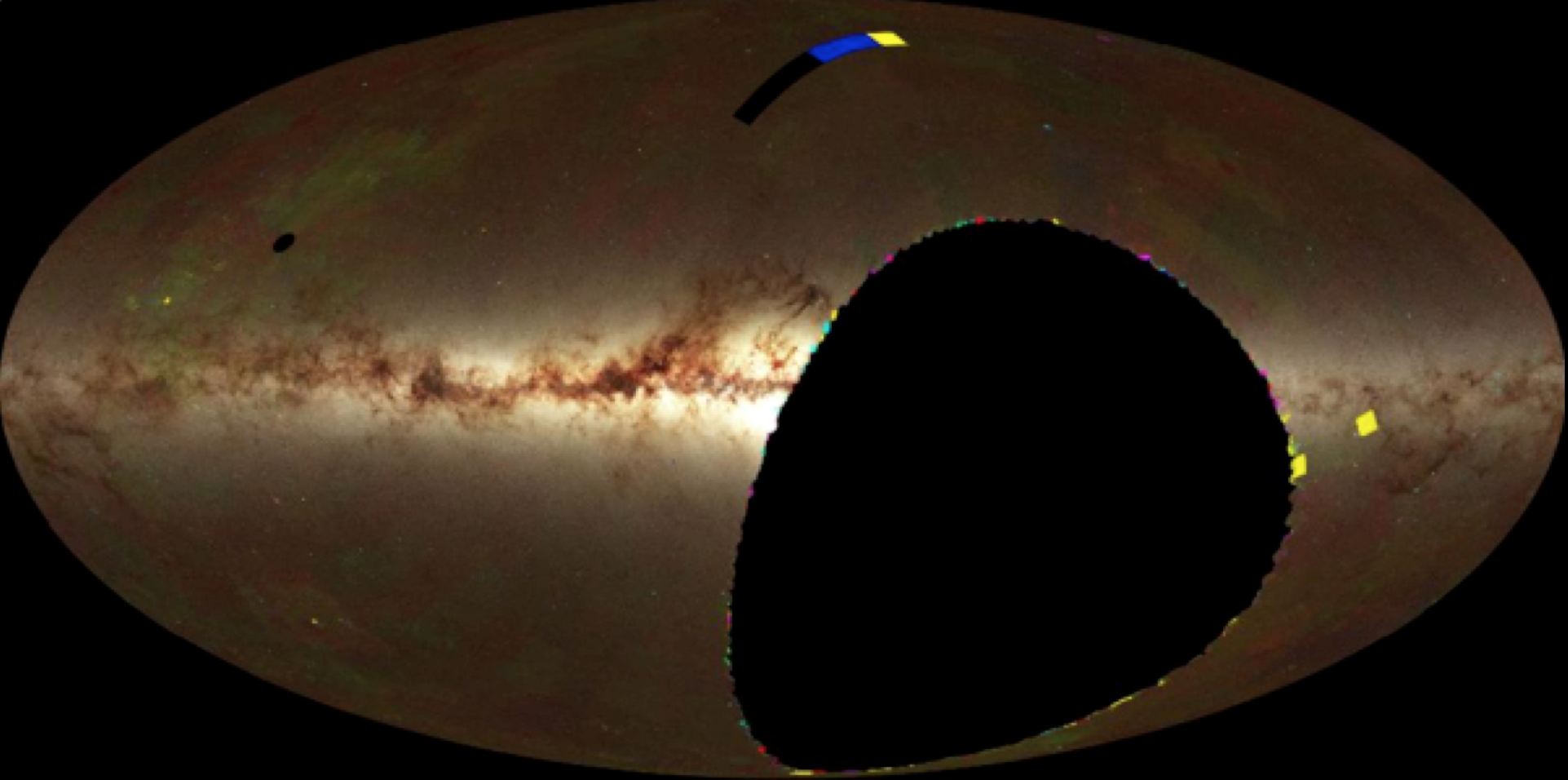}
\caption{The Pan-STARRS PS1 $3\pi$ survey. The unsurveyed area has Dec$<-30$deg.}
\label{fig:ps1_3pi}       
\end{figure}

\section{Pan-STARRS $3\pi$ Survey}

The Pan-STARRS $3\pi$ survey (Tonry et al. 2012) is an ambitious survey
aimed at covering 3/4 of the sky in the $grizy$ optical-NIR bands (see
Fig. 8). It uses a 1.8-m, $f/4.4$ telescope at Haleakala Observatory on
Maui, Hawaii with an $\approx 7$deg$^2$ field-of-view. The $3\pi$ survey
was done with a cadence of 6 epochs with each epoch consisting of a pair
of exposures taken $\approx 25$mins apart. Individual  exposure times
vary between 30-45s skewed towards the bluer bands. The detectors are
OTA arrays which have high red sensitivity and pixel widths of $0.''26$.
Seeing is typically $1.''1-1.''3$. It is interesting to compare light
grasp $+$ throughput with VST ATLAS. ATLAS has a $2.1\times$ bigger
mirror area and better throughput in $gri$. We estimate that PS1\footnote{PS1 
is the first of 4 planned Pan-STARRS telescopes.} has
$\approx1.5\times$ higher throughput in the $z$ band and $2\times$ less
throughput in the $g$ band.  We make a comparison of ATLAS exposure
times, median seeing and magnitude limits with Pan-STARRS $3\pi$ (and
KiDS and DES) in Table \ref{tb:exp_time}). The total exposure time of
PS1 is $5.2\times$ longer in $g$ and $4\times$ longer in $i$.  However,
the seeing is 20\% worse in all bands and sky brightnesses are
approximately the same. Point source magnitude limits should therefore
be similar in $gr$, $\approx0.5$mag fainter in $i$ and $\approx1$mag
fainter in $z$ (see Table \ref{tb:exp_time}). Similar conclusions apply
to $gri$  for the SDSS-PanSTARRS comparison but the PanSTARRS  advantage
over SDSS  in $z$ is now $\approx 1.7$mag for stars (Shanks et al. 2015).

\begin{table}
\centering
\begin{tabular}{ccccc}
\hline\hline
Survey/Band & $g$ & $r$ & $i$ & $z$ \\
\hline\hline

ATLAS Exposure  & $2\times50$s & $2\times45$s & $2\times45$s & $2\times45$s \\
Mean PS1 Exposure & $8.4\times43$s & $8.4\times40$s & $8.8\times45$s & $9.7\times30$s \\
DES Y1 Exposure & $4\times90$s & $4\times90$s & $4\times90$s & $4\times90$s \\
KiDS Exposure   & $900$s       & $1800$s      & $1080$s      & $500$s \\
DES Y5 Exposure & $10\times90$s& $10\times90$s& $10\times90$s& $10\times90$s \\
\hline
ATLAS Seeing    & $0.''95$ & $0.''90$ & $0.''81$ & $0.''84$ \\
PS1 Seeing      & $1.''33$ & $1.''19$ & $1.''13$ & $1.''08$ \\
KiDS Seeing     & $0.''8$  & $0.''7$  & $0.''75$ & $0.''8$ \\
DES Seeing      & $1.''0$  & $0.''94$ & $0.''94$ & $0.''94$ \\ 
\hline
ATLAS Mag Lim   & $23.14$  & $22.67$  & $21.99$  & $20.87$ \\ 
PS1 Mag Lim     & $23.05$  & $22.85$  & $22.45$  & $21.85$ \\ 
DES Y1 Mag Lim  & $25.0$   & $24.5$   & $24.1$   & $23.3$ \\ 
KiDS Mag Lim   & $25.0$   & $24.8$   & $23.8$   & ($23.1$) \\ 
DES Y5 Mag Lim & $25.5$   & $25.0$   & $24.6$   & $23.8$ \\ 

\hline
\end{tabular}
\caption[]{ATLAS-PanSTARRS PS1 $3\pi$-KiDS-DES comparison. ATLAS median
seeing for ESO A,B classified tiles. ATLAS Mag Lim  corresponds to the
median $5\sigma$ magnitude detection limit for stars as measured in a
$1''$ radius aperture. PS1 Mag Lim is the measured $5\sigma$ magnitude
limit inside a $3''$ diameter aperture. KiDS and DES Mag Lims are
$5\sigma$ limits for stars aperture corrected in a $2''$ diameter
aperture. Sky brightness is measured in ABmag/arcsec$^2$. All magnitudes
are quoted in the AB system. DES Y1 and Y5 refers to the DES Year 1 and Year 5 
exposures and magnitude limits.
}
 \label{tb:exp_time}
\end{table}


\section{Dark Energy Survey}

The Dark Energy Survey (DES) is aimed at KiDS depths over ATLAS areas.
It uses the Dark Energy Camera (Flaugher et al 2015) installed at the prime focus of the CTIO
4-m Blanco Telescope. It is aimed at covering 5000deg$^2$ over a period
of 5 years starting in August 2013. It aims to do a pass with
$4\times90$s exposures in $grizy$ for the first 2 years increasing this
to $10\times90$s by the end of Year 5. DECam has high throughput
especially in $i$ and $z$ due to its deep depletion CCDs. The DECam
pixel size is $0.''263$. The quality of its imaging can be seen in Fig. 9 which 
compares DES and HST images of gravitationally lensed arcs in  a galaxy cluster.

\begin{figure}[t]
\includegraphics[scale=.175]{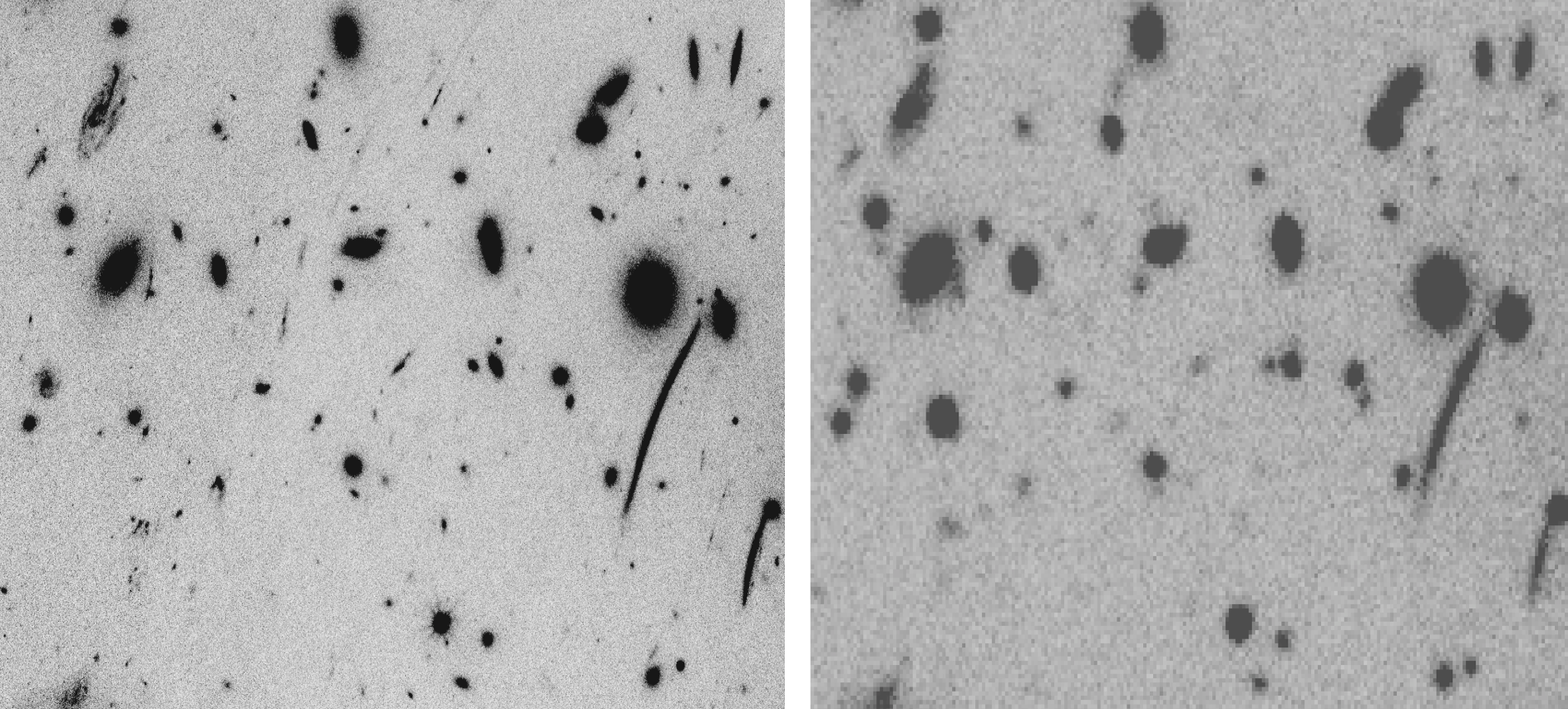}
%
%
\caption{Left: HST F435W image of galaxy cluster RXJ2248 from the CLASH collaboration. 
Right: DES $riz$ image of the same cluster.}
\label{fig:des_cluster}       
\end{figure}

The aims of DES are mostly cosmological to determine the dark energy
equation of state by a variety of probes including cluster counts, weak
lensing and large-scale structure. They are also covering 30deg$^2$ in
time domain mode to search for supernovae. But it is the wide field
aspect of DES that is our focus here. The cluster counts include
cross-correlation with the South Pole Telescope CMB SZ measurements (see Fig. 10).
Much will be learned about the astrophysics of cluster gas as well as
cosmology. The large-scale structure aims include galaxy and quasar
power spectra  at the largest scales partly to constrain neutrino
masses. Here there will be a need for accurate photo-z but these will
only be available out to $z\approx1$ given the lack of deeper NIR
photometry beyond $Y$. Clearly the VHS $JHK_s$ photometry will provide
opportunities for very high redshift quasar z-dropouts at $z>7$ but
there may be a future need for VISTA to provide $JHK_s$ imaging to VIKING
depths to fully exploit DES $grizy$ over the full survey area.

\begin{figure}[t]
\includegraphics[scale=.35]{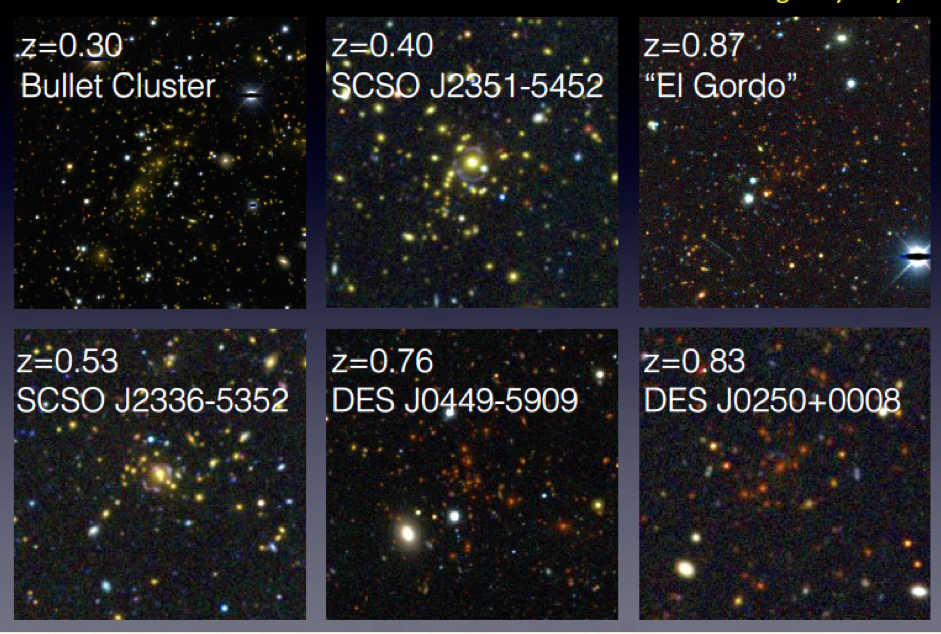}
\caption{DES mages of lensing clusters out to redshift, $z\approx0.9$ (E. Rykoff et al. priv. comm.).}
\label{fig:des_clusters}       
\end{figure}

For DES, the main target must be  weak-lensing shear. DES certainly
provides enough $r$-band exposure to reach $r\approx25$ at 5$\sigma$ after
5 years. The other two necessary components are the consistency of the
PSF over the full focal plane and the delivered seeing. Fig. 11 shows
that it is possible for DES to provide  median $0.''94$ seeing in the
$r$ band, close to its $0.''9$ specification for weak lensing. However,
this is mainly produced by sub-selecting the best seeing for $r$. There
also seems to be a lower limit to DECam seeing of $0.''75$ in the $r$
band. On the issue of the uniformity of the PSF there is less
information available, although weak lensing maps have already been
produced using DES data.

\begin{figure}[t]
\centering{
\includegraphics[scale=.175]{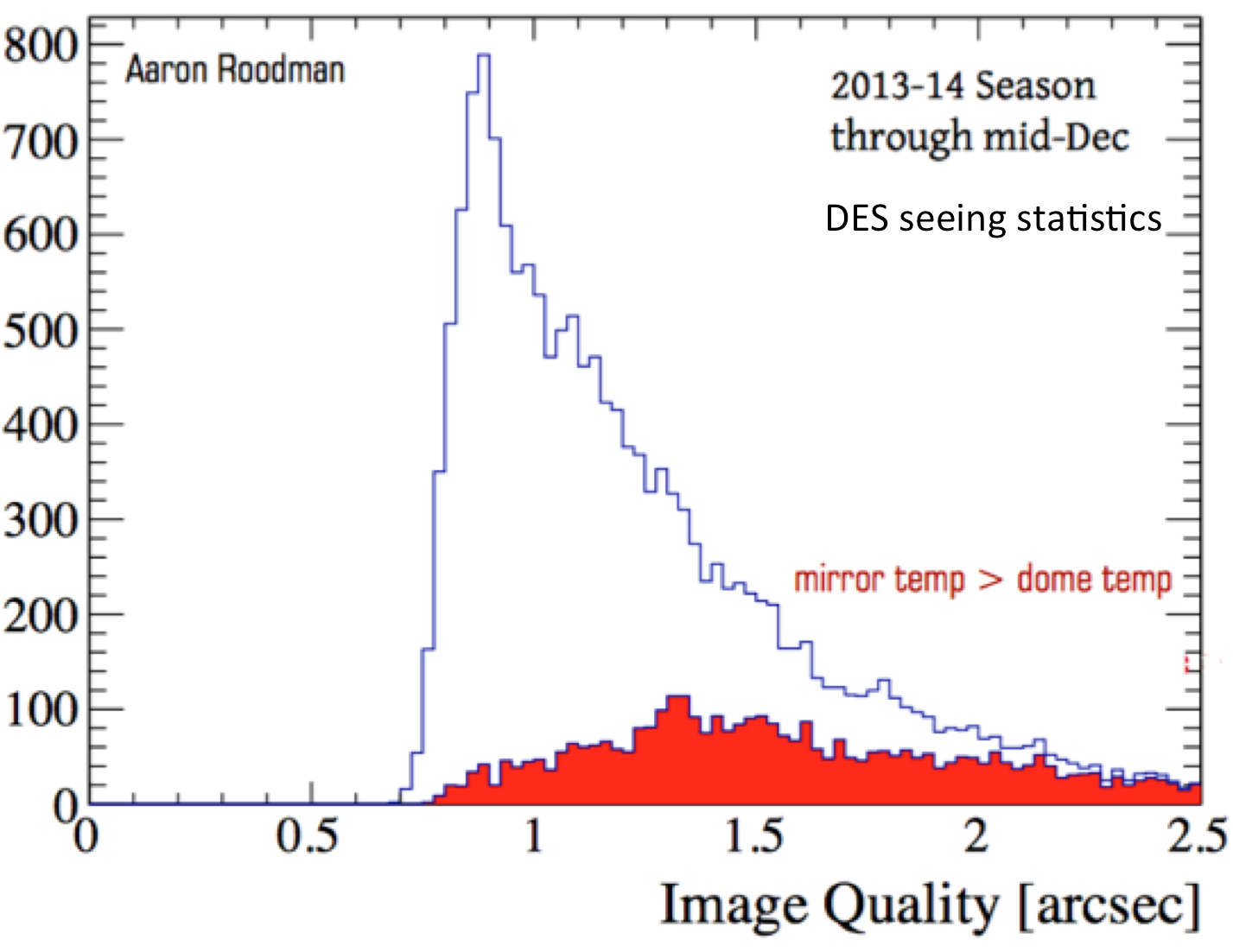}
\caption{DES seeing statistics from  Year 1 (A. Roodman priv.
comm.). The median (sub-selected) seeing in the $r$-band is $0.''94$ FWHM compared to
$0.''9$ specification. The red histogram demonstrates the importance of
maintaining the mirror temperature below the dome temperature.}
}
\label{fig:roodman}       
\end{figure}

\section{DES versus other surveys}

Clearly other surveys are now playing David to the Goliath survey that
is DES. In terms of DES versus ATLAS, DES reaches further South on the
SGC side of the sky than ATLAS whereas ATLAS also covers the NGC side.
The 5th band for ATLAS is the $u$ band to $u=22.4$ (including Chilean
ATLAS $u$ extension) whereas the 5th band for  DES is $y$ with the
$5\sigma$ limit $y=22.7$ in Year 5. As can be seen from Table 2 ATLAS
median seeing is slightly better in all bands than DES. Pan-STARRs
$3\pi$ also has its 5th band in $y$ rather than $u$, reaching $y=20.85$
at $5\sigma$. Again its seeing is worse than both DES and ATLAS. KIDS is
most  competitive with DES in the $r$ band where KiDS'  seeing is
significantly better (see Fig. 12).   DES Y5 $r$ magnitude limit is
0.2mag fainter. In the other bands, DES Y5 will be significantly
(0.5-0.8mag) fainter than the KiDS limits. VIKING, however, is well
matched to KiDS and VHS is not so well matched to DES except for the
purpose of high redshift quasar searches. For many other purposes VHS is
a better match to ATLAS. The same could be said for WISE - it is a
better match to SDSS depth surveys such as ATLAS than it is to KiDS or
DES.

SDSS has also been a great success because of its access to Luminous Red
Galaxies extending out to $z\approx0.75$. This can now be extended to
$z\approx1$ by incorporating  WISE W1 and W2 data into ATLAS. By
$z\approx1.3$ the 4000A break has redshifted out of the optical bands
and $z>1.3$ LRGs are likely too faint even for UKIDSS LAS and VHS.
Emission line galaxies out to $z\approx0.7$ can be selected at ATLAS
depth (Bielby et al 2012) but it is undeniable that deeper surveys such
as DES can in principle probe out to $z\approx2$. But the lack of
matched NIR data will make it difficult to measure photo-z past the VHS
limit. However, the lack of a deep $u$ band makes it hard to look for
$2<z<3.5$ dropouts. The $3.5<z<4.5$ $g$ dropouts are a possibility but
even the DES 5-year limits will make it tough to detect these.

The basic  advantage of SDSS depth in the optical is that the galaxy
number counts turn over within the SDSS range and most of the fainter
galaxies are also intrinsically faint, so the majority of galaxies are
probing the same volume as those at brighter limits. Of course, the
brightest of the more distant galaxies are there but until $g$ dropouts
become available at $z\approx4$, these will remain expensive to identify
even for a survey like DES. The effective volume of the lower redshift
is optimised where the galaxies reach the knee in the luminosity
function at M$^*$. At this point correlation functions and power spectra
produce their smallest errors in the most cost-effective  exposure time.

\begin{figure}[t]
\centering{
\includegraphics[scale=.25]{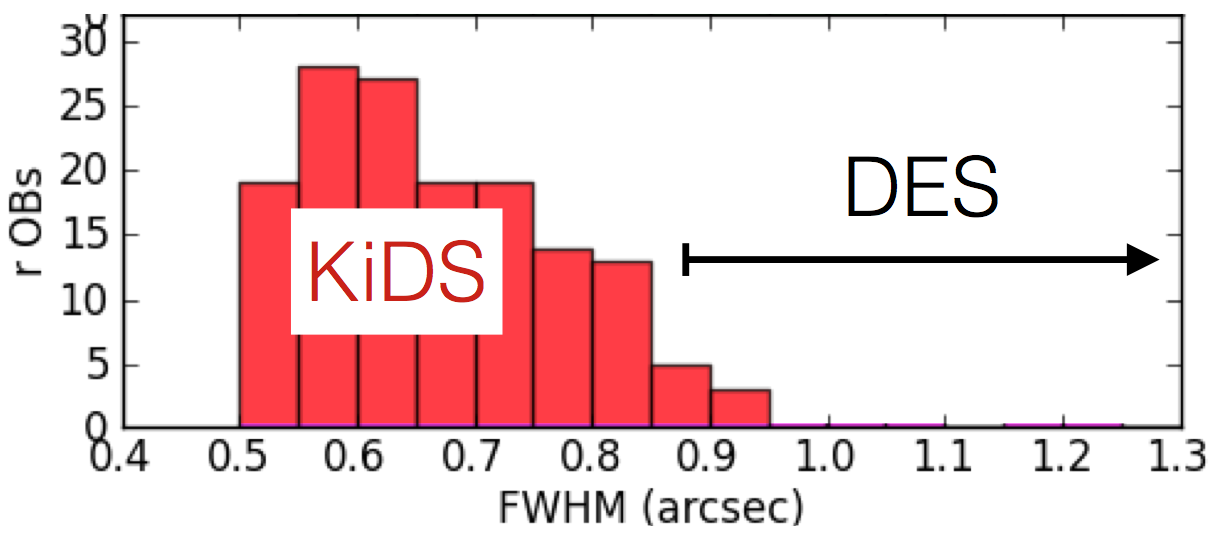}
\caption{KiDS seeing statistics compared  to the DES Year 1 range from
Fig. 10 (K. Kuijken, priv. comm.)}
}
\label{fig:4}       
\end{figure}

The great overpowering advantage that the depth of DES allows is of
course for cosmological weak-lensing studies. The image quality is
$0.''94$ and this is judged adequate for this purpose. The main
competition here is from KiDS where the median seeing is $0.''7$ a 25\%
improvement on DES. The KiDS magnitude limit in $r$ is 24.8mag which is
lies between the  DES 1-2 year  and 5 year limits. Of course,
weak-lensing demands uniform PSF's across the field-of-view and this has
already been demonstrated for KiDS and this is still under scrutiny for
DES. Speed of survey is also slower for KiDS than DES given its
$\approx3\times$ bigger field and $\approx2.4\times$ bigger mirror area.
The KiDS SGC area is half covered by DES and so this should make for
interesting lensing comparisons with DES. In this area at least, DES will
ultimately be able to incorporate the VIKING $JK_s$ data for photo-z.

\section{Conclusions}

\subsection{No access to DES? Stay calm and don't panic!}

So what can be done if you are at an institute with no access to DES?
The message is to stay calm - in the Southern sky  at least  the ESO
Public Surveys can help. Much of the low lying fruit can be accessed
with SDSS depth surveys such as ATLAS and then there is KiDS, VIKING and
VHS too. Over wider areas  the PanSTARRS $3\pi$ survey will soon become
available and its $\approx70$deg$^2$ of Medium Deep Survey
fields will also be available. For ATLAS there is the pleasing match of the WISE
Mid-IR surveys to its depth, now being exploited for LRG and high-z quasar
studies. The prime science area for DES is weak lensing but even here
KiDS can compete with its combination of similar $r$-band depth and
$\approx25$\% better seeing.

In the future, the 8-m LSST (Tyson, 2005) will reach  an AB depth of 26.1-27.5 in
$ugrizy$ over the whole Southern sky. For synoptic surveys, LSST will
have an unparalleled reach but meanwhile  there is much work that  can be done now
in PS1 MDS and the overlap areas between PS1 $3\pi$, ATLAS, SDSS, KiDS
and DES, not to mention VHS and VIKING. 

\subsection{But no access to spectroscopic follow-up? }

For LSST photometric redshifts are hugely important. Indeed, the LSST
`black book' barely makes mention of spectroscopic follow-up. Generally,
the Southern sky looks weakest compared to the Northern Hemisphere in
terms of spectroscopic follow-up. There is AAT 2dF $+$ HERMES and there
will be MOONS but the latter has a small  $\approx25'$ field similar to
Gemini FMOS but with more fibres ($\approx1000$). VISTA 4MOST with
$\approx1000$ fibres over 3deg$^2$ should be excellent for ATLAS
follow-up but perhaps needs more fibre numbers to match the high sky
densities of galaxies and stars available from the deeper surveys. There
have been more ambitious designs made such as NG2dF and  VXMS to make
multislit spectrographs to cover $>3$deg$^2$ but these have generally
fallen foul of the current need for high resolution ($R>10000$)
spectroscopy for Galactic Archaeology (coupled to GAIA astrometry),
which is done better with fibres. But there remains possibilities at 2dF
and at DECam to build extreme multislit spectrographs where
$\approx10000$ galaxy redshifts can be measured simultaneously.
Combining photo-z information from the deep imaging surveys with low S/N
spectroscopic redshifts, reduce the redshift error from
$\pm10000$kms$^{-1}$ to $\pm100$kms$^{-1}$ is also  a possibility that
is worth investigating. The main aim of these surveys would be to
measure redshift-space distortions combined with lensing tomography.

There may be a contribution to spectroscopic surveys that VST can make
once its current imaging surveys are completed. Even extreme multislit
spectrographs may miss interesting objects because  of the need to
target previously imaged galaxies and stars to manufacture the slit
mask. However slitless spectroscopy could offer a potentially exciting
route for a telescope like VST. An objective prism is infeasible because
of the large mirror diameter. Instead, it has been suggested that a
grens for VST that might be easily accommodated in the filter wheel
could offer low dispersion spectroscopy over the full 1 degree field (R.
Content, 2014, priv. comm.). There would be a price  to pay in terms of
sky noise but previous experience at UKST and CFHT suggest that 30 min
exposures could produce good emission line spectra for  $\approx100$
$g<22$ mag quasars in a typical VST field. It  may also be  possible
simultaneously to measure redshifts for at least  several hundred strong
emission- and absorption-line galaxies.  But for slitless the main
advantage is that we get to observe ALL  the spectra with no need for
pre-selection for fibres or masks, with immediate advantage for future searches
for gravitational wave source counterparts etc.

\subsection{Ground-based image competition for EUCLID?} Given the
excellent seeing that VST has experienced, for the final perspective, it
may also be worth thinking about improving its optics to see if even
better seeing could be accessed. It might not take too much of an
improvement to make VST competitive with EUCLID, for example, at least
over small areas of a few hundred square degrees with `sub-selected'
ground-based seeing. Even if this proved impossible with VST, it would
make a good proving ground for a more sophisticated telescope of similar
size to see what are the limits of what can be achieved by a
ground-based telescope on an excellent site such as Paranal. Recall that
this  goal of using active {\it and} adaptive optics over a wide field 
was already the ambitious original aim of the PanSTARRS project. Maybe
this idea came too early for the technology and perhaps its time 
will come again?

\begin{acknowledgement}
I should like to thank N. Metcalfe and B. Chehade (Durham Univ.),
M.J.Irwin and E. Gonzalez-Solares (CASU Cambridge) and R.G. Mann, M.A.
Read and colleagues from WFAU, Edinburgh for all their help with the VST
ATLAS survey. We are also deeply grateful to the OmegaCAM and VST teams
and the   ESO VST observers whose dedication makes the ESO VST surveys
possible. I should also like to thank K Kuijken (Leiden), A.C. Edge
(Durham), O. Lahav (UCL) and S. Smartt (QUB) for allowing me to present
their images illustrating highlights from the KiDS, VIKING, DES and
PanSTARRS surveys. Finally I should like to thank the Scientific Organising
Committee for inviting me to give this review at the Universe of Digital Sky 
Surveys meeting in honour of the 70th birthday of Prof. Massimo Capaccioli.
\end{acknowledgement}
%

%
%
%

\def \mnras {MNRAS}
\def \apj {ApJ}
\def \aj {Astr.J}
\def \aap {Astr. Astrophys.}
\def \procspie {Proc. SPIE}

\end{document}